\documentclass[aps,prl,twocolumn,showpacs]{revtex4-1}
\usepackage{epsfig}
\usepackage{bm}
\usepackage{color}
\usepackage{ulem}
\newcommand{\be}{\begin{eqnarray}}
\newcommand{\ee}{\end{eqnarray}}
\newcommand{\ba}{\begin{array}}
\newcommand{\ea}{\end{array}}
\newcommand{\bmat}{\left(\begin{array}}
\newcommand{\emat}{\end{array}\right)}
\newcommand{\no}{\nonumber}
\newcommand{\Tr}{\mbox{Tr}\,}

\begin{document}
\title{Comment on ``Energy-time uncertainty relation for driven quantum systems'' and ``Quantum Speed Limit for Non-Markovian Dynamics''}
\author{Manaka Okuyama$^1$}
\author{Ryo Takahashi$^1$}
\author{Masayuki Ohzeki$^2$}%
\affiliation{%
$^1$Department of Physics, Tokyo Institute of Technology, Oh-okayama, Meguro-ku, Tokyo 152-8551, Japan
}
\affiliation{%
$^2$Graduate School of Information Sciences, Tohoku University, Sendai 980-8579, Japan
}

\date{\today}

\begin{abstract} 
Deffner and Lutz [J. Phys. A {\bf 46}, 335302 (2013) and Phys. Rev. Lett. {\bf 111}, 010402 (2013).] extended the Mandelstam-Tamm bound and the Margolus-Levitin bound to time-dependent and non-Markovian systems, respectively.
Although the derivation of the Mandelstam-Tamm bound is correct, we point out that thier analysis of the Margolus-Levitin bound is incorrect. 
The Margolus-Levitin bound has not yet been established in time-dependent quantum systems, except for the adiabatic case.



\end{abstract}

\maketitle

The quantum speed limit (QSL) gives the fundamental speed limit to quantum time evolution. 
In time-independent quantum systems, the minimal evolution time $\tau_{\text{QSL}}$ needed for the state to rotate orthogonally is given by
\be
\tau &\ge& \max \left\{ \frac{\pi \hbar}{2\Delta E} ,  \frac{\pi \hbar}{2  \left(E-E_0 \right)}  \right\} ,
\ee
where $\Delta E$, $E$ and $E_0$ are the energy variance, mean energy and ground-state energy, respectively. 
The first bound is called  the Mandelstam-Tamm (MT) bound \cite{MT} and the second bound is called  the Margolus-Levitin (ML) bound \cite{ML}.
We emphasize that the MT and ML bounds are characterized by the energy variance and mean energy, respectively.

Recently, Deffner and Lutz derived the two ML bounds in time-dependent systems \cite{DL}.
Furthermore, in Ref. \cite{DL2}, they derived the MT bound  and the ML bound for non-Markovian dynamics. 
As a result, they concluded that the ML bound is tighter than the MT bound in non-Markovian systems. 

In this comment, we point out the following:
(i) The derivation of one ML bound for unitary dynamics in Ref. \cite{DL} is incorrect.
(ii) Another ML bound for unitary dynamics in Ref. \cite{DL} has no physical meaning.
(iii) The derivation of the ML bound for unitary dynamics in Ref. \cite{DL2} is incorrect.
(iv) The inequalities for non-Markovian dynamics in Ref. \cite{DL2} cannot be regarded as the ML bound and  has no physical meaning.
(v) The ML bound has not yet been established in time-dependent quantum systems, except for the adiabatic case \cite{AA}.

In Ref. \cite{DL}, the authors used the following relation
\be
|\langle\psi_0|\psi_\tau\rangle|=|\langle\psi_0| U_\tau|\psi_0\rangle|=\left| \sum_n |\langle \psi_0|n \rangle|^2 \exp(-iJ_n) \right|
\label{t-order} ,
\no\\
\ee
where they defined that $U_\tau$ denotes the time evolution operator and $\{ |n\rangle \}$ is the set of its instantaneous eigen states, with 
\be
\frac{1}{\hbar} \int dt_0^\tau H_t |n\rangle \equiv J_\tau|n\rangle =J_n |n\rangle \label{t-order2}.
\ee
Using Eq. (\ref{t-order}), the authors obtained the ML bound for time-dependent closed systems
\be
\tau \ge \frac{\hbar}{E_\tau}  \mathcal{L}(\psi_0,\psi_\tau) \label{ML-time} ,
\ee
where $E_\tau=(1/\tau)\int_0^\tau dt \left| \langle \psi_0| H_t|\psi_0 \rangle \right|$ and $\mathcal{L}(\psi,\psi_\tau)=\arccos(|\langle \psi_0|\psi_\tau \rangle |)$.

However, Eq. (\ref{t-order}) does not hold clearly.
The authors identified $\exp(-(i/\hbar) \int_0^\tau dt H_t)$ with $U_\tau$ and ignored the time ordered product of $U_\tau$, which is never justified.	
In order to correctly realize their idea, we must use the Magnus expansion \cite{Magnus}:
\be
U_\tau&=&\exp\left(- \frac{i}{\hbar}  \Omega_\tau \right),
\\
\Omega_\tau &=& \int_0^\tau dt_1 H_{t_1} -\frac{i }{2\hbar} \int_0^\tau dt_1 \int_0^{t_1} dt_2 \left[ H_{t_1}, H_{t_2} \right] 
\no\\
&&+\cdots ,
\\
\Omega_\tau |n'\rangle &=&J_n'|n'\rangle ,
\ee
where $|n'\rangle$ is the set of  instantaneous eigenstates of $\Omega_\tau$.
Then, we can identify $\exp\left(- (i/\hbar) \Omega_\tau \right)$ with  $U_\tau$ and Eq. (\ref{ML-time}) is modified to 
\be
\tau \ge \frac{\hbar}{ \frac{1}{\tau}\left| \langle \psi_0|\Omega_\tau|\psi_0 \rangle \right|}  \mathcal{L}(\psi_0,\psi_\tau) \label{ML-time2} .
\ee
Although the derivation of Eq. (\ref{ML-time2}) is correct, it is a formidable task to estimate the value of $\left| \langle \psi_0|\Omega_\tau|\psi_0 \rangle \right|$ via $H_t$ in general.

In addition, the authors derived also another ML bound in appendix of Ref. \cite{DL} 
\be
\tau \ge \frac{4\hbar}{\pi^2 \bar{E}_\tau}  \mathcal{L}^2(\psi_0,\psi_\tau) \label{ML-time3} ,
\ee
where $\bar{E}_\tau$ is given by $(1/\tau) \int_0^\tau dt \left| \langle \psi_0|H_t |\psi_t\rangle \right|$.
Although the derivation of Eq. (\ref{ML-time3}) is correct, the value of $\langle \psi_0|H_t |\psi_t\rangle$ cannot be limited only from the eigenvalues of $H_t$.
Therefore, Eqs. (\ref{ML-time2}) and (\ref{ML-time3}) are mathematically correct but have no physical meaning. The authors failed to obtain the meaningful ML bound for time-dependent closed systems in Ref. \cite{DL}. 

In Ref. \cite{DL2}, the authors first considered time-dependent closed systems and used 
\be
\text{tr}\left\{ |H_t \rho_t |\right\} =\langle H_t \rangle, \label{tr}
\ee
where $\text{tr}$ means the trace norm and $\rho_t= |\psi_t \rangle \langle \psi_t |$.
Using this relation, the authors obtained the ML bound for time-dependent closed systems
\be
\tau \ge \frac{\hbar}{2E_\tau'} \sin^2 \left( \mathcal{L}(\rho,\rho_\tau)\right) \label{ML-close},
\ee
where $E_\tau'=(1/\tau)\int_0^\tau dt \left| \langle \psi_t| H_t|\psi_t \rangle \right|$ and $\mathcal{L}(\rho,\rho_\tau)=\arccos(\sqrt{ {\rm tr}(\rho \rho_\tau)})$.

However, Eq. (\ref{tr}) does not hold. 
Correctly, $\Tr\left\{ |H_t \rho_t |\right\} $ is evaluated  as $ \sqrt{\langle \psi_t|H_t^2 | \psi_t\rangle}$ ,
and Eq. (\ref{ML-close}) is modified to 
\be
\tau \ge \frac{\hbar}{ \frac{2}{\tau} \int_0^\tau dt  \sqrt{\langle \psi_t|H_t^2 | \psi_t\rangle} 
} \sin^2 \left( \mathcal{L}(\rho,\rho_\tau)\right) . \label{ML-close2}
\ee
Using $\sqrt{\langle \psi_t|H_t^2 | \psi_t\rangle} \ge \sqrt{\langle \psi_t|H_t^2 | \psi_t\rangle-(\langle \psi_t|H_t | \psi_t\rangle)^2} $, we immediately find that Eq. (\ref{ML-close2}) is looser than the MT bound (which was also obtained in Ref. \cite{DL2}) 
\be
\tau \ge \frac{\hbar}{ \frac{\sqrt{2}}{\tau} \int_0^\tau dt \Delta E_t} \sin^2 \left( \mathcal{L}(\rho,\rho_\tau)\right), \label{MT-close}
\ee
where $\Delta E_t =\sqrt{\langle \psi_t|H_t^2 | \psi_t\rangle-(\langle \psi_t|H_t | \psi_t\rangle)^2}$.

Therefore, we conclude that Eq. (\ref{ML-close}) does not hold in time-dependent closed systems and Eq. (\ref{ML-close2}) regarded as the ML bound in Ref. \cite{DL2} gives the  looser bound than the MT bound (\ref{MT-close}).

In the latter part of Ref. \cite{DL2}, the authors considered non-Markovian systems and obtained the following inequalities 
\be
\tau &\ge& \max \left\{  \frac{1}{\Lambda_\tau^{\text{op}}} , \frac{1}{\Lambda_\tau^{\text{tr}}} , \frac{1}{\Lambda_\tau^{\text{hs}}} \right\}
\sin^2\left( \mathcal{L}(\rho, \rho_\tau) \right)  \label{nonM-ML-MT} ,
\ee
where $\Lambda_\tau^{\rm op, tr, hs}=(1/\tau)\int_0^\tau dt || \dot{\rho}_t||_{\rm op, tr, hs}$, $|| A||_{\text{op}}={\sigma_1} $, $|| A||_{\text{tr}}=\sum_i \sigma_i $, $|| A||_{\text{hs}}=\sqrt{\sum_i \sigma_i^2} $, $\sigma_i$ are the singular values of $A$ and $\sigma_1$ is the largest singular value of $A$.
Furthermore,  using the trace inequality $|| A||_{\text{op}} \le || A||_{\text{hs}} \le || A||_{\text{tr}}$, Eq. (\ref{nonM-ML-MT}) is deduced to 
\be
\tau \ge\frac{1}{\Lambda_\tau^{\text{op}}}\sin^2\left( \mathcal{L}(\rho, \rho_\tau) \right) \label{nonM-ML}  .
\ee
The authors  regarded   $1/\Lambda_\tau^{\text{op}}$ and $1/\Lambda_\tau^{\text{tr}}$ as the ML type bounds and $1/\Lambda_\tau^{\text{hs}}$ as the MT type bound. 
Therefore, they concluded that the ML type bound is the sharpest bound in non-Markovian systems.

However, when we consider unitary dynamics $\dot{\rho}_t =(1/\hbar) \left[H_t, \rho_t \right]$,
 we cannot regard $1/\Lambda_\tau^{\text{op}}$ and $1/\Lambda_\tau^{\text{tr}}$ as the ML type bounds because $|| \dot{\rho}_t||_{\text{op}}$ and $|| \dot{\rho}_t||_{\text{tr}}$ are different from the mean energy (while $|| \dot{\rho}_t||_{\text{hs}}$ is equal to $\sqrt{2} \Delta E_t /\hbar$ and $1/\Lambda_\tau^{\text{hs}}$ is deduced to Eq. (\ref{MT-close}) exactly, that is, the MT bound.).
Their physical meaning is unknown ever for unitary dynamics and, therefore, Eq. (\ref{nonM-ML})  is mathematically correct but has no physical meaning for non-Markovian dynamics. The ML bound  has not been found in non-Markovian systems so far.

In summary, the ML bound is limited only to time-independent systems and has not yet been established in time-dependent systems except for the adiabatic case \cite{AA}.  
The derivation of the ML bound is based on spectrum expansion \cite{ML}  and, when we extend it straightforwardly, we obtain Eq. (\ref{ML-time2}) which makes no sense physically.
In addition, we mention that, for the classical Liouville equation, the classical ML-type bound is looser than the classical MT-type bound even in time-independent systems \cite{OO}.
These results might imply that the  ML bound is  a peculiar phenomenon to time-independent (or adiabatic) systems and not a universal property in time evolution.



\section*{Acknowledgments}
M. Okuyama was supported by JSPS KAKENHI Grant No. 17J10198.
M. Ohzeki was supported by ImPACT Program of Council for Science, Technology and Innovation (Cabinet Office, Government of Japan) and JSPS KAKENHI No. 16K13849, No. 16H04382 and the Inamori Foundation.



\section*{References}
\begin{thebibliography}{99}

\bibitem{MT}
L. Mandelstam and I. G. Tamm,
{ The uncertainty relation between energy and time in nonrelativistic quantum mechanics}, 
J. Phys. (Moscow) {\bf 9}, 249 (1945).


\bibitem{ML}
N. Margolus and L. B. Levitin,
{ The maximum speed of dynamical evolution}, 
Physica D {\bf 120}, 188 (1998). 

\bibitem{DL}
S. Deffner and E. Lutz,
{ Energy-time uncertainty relation for driven quantum systems}, 
J. Phys. A {\bf 46}, 335302 (2013).

\bibitem{DL2}
S. Deffner and E. Lutz,
{ Quantum Speed Limit for Non-Markovian Dynamics},
Phys. Rev. Lett. {\bf 111}, 010402 (2013).

\bibitem{AA}
M. Andrecut and M. K. Ali,
{ The adiabatic analogue of the Margolus-Levitin theorem},
J. Phys. A {\bf 37}, L157 (2004)

\bibitem{Magnus}
W. Magnus, 
{ On the exponential solution of differential equations for a linear operator}, 
Commun. Pure Appl. Math. {\bf VII}, 649 (1954).

\bibitem{OO}
M. Okuyama and M. Ohzeki,
{ Quantum Speed Limit is Not Quantum},
arXiv:1710.03498 (2017).


\end{thebibliography}
\end{document}